\documentclass[11pt]{article}
\usepackage{hyperref}
\usepackage{graphicx}
\def\bea{\begin{eqnarray}}
\def\eea{\end{eqnarray}}
\def\beq{\begin{equation}}
\def\eeq{\end{equation}}
\def\nn{\nonumber}

\def\a{\alpha}
\def\b{\beta}
\def\e{\epsilon}
\def\r{\rho}

\def\th{\theta}

\def\D{\Delta}

\def\d{\delta}

\def\tb{\bar{t}}
\def\phib{\bar{\phi}}

\begin{document}

\begin{center} {\Large \bf
Where is the extremal Kerr ISCO?}
\end{center}

\vskip 5mm
\begin{center} \large
{{Ted Jacobson
}}
\end{center}

\vskip  0.5 cm
{\centerline{\it Center for Fundamental Physics}}
{\centerline{\it Department of Physics, University of Maryland}}
{\centerline{\it College Park, MD 20742-4111, USA}}

\vskip 1cm

\begin{abstract}
Although the circular photon orbit and ISCO for 
the Kerr black hole remain distinct
from each other and from the horizon in the extremal spin limit
on a constant Boyer-Lindquist time slice, on a horizon-crossing
slice they both coincide with the null generators of the horizon. 
\end{abstract}

\section{Introduction}
In a classic paper on the physics of Kerr black holes, Bardeen, Press
and Teukolsky\cite{Bardeen:1972fi} mentioned in passing a peculiar 
feature of the prograde circular 
photon and innermost stable (ISCO) orbits in the extremal limit: 
although the Boyer-Lindquist (BL) radial 
coordinates of these orbits both coincide with that of the horizon, their
locations on a BL time slice are both distinct from the horizon and from 
each other. The photon orbit lies at a finite distance from the horizon, 
while the ISCO lies at an infinite distance from the horizon, 
and also at an infinite distance 
from any point with a greater value of the radial coordinate.\footnote{In
Ref.~\cite{Bardeen:1972fi}
the marginally bound orbit was also discussed.
In the extremal limit it lies a finite distance outside the photon orbit.}
This strange behavior, which will be explained in detail below, 
is tied to the fact that a BL time slice ends on the bifurcation surface 
(the intersection of the future and past horizons), and 
in the extremal limit the distance to the bifurcation surface becomes infinite. 
The geometry of the situation is illustrated in Fig.\ 1, which shows the
causal structure of the equatorial plane of the Kerr spacetime in the region
exterior to the future and past horizons.
\begin{figure}[h]
\begin{center}
\includegraphics[width=2.5in]{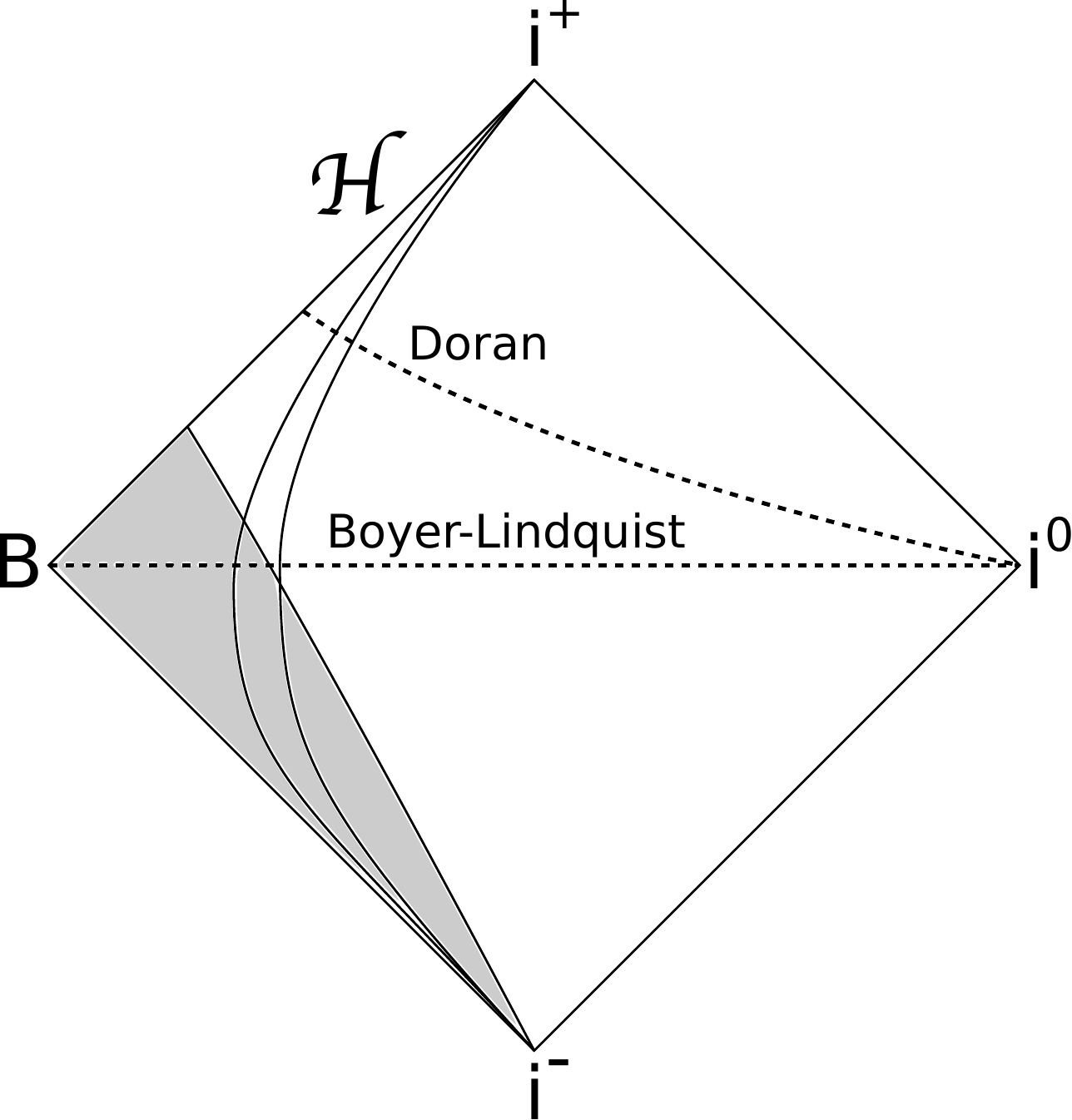}
\end{center}
\caption{\small
Carter-Penrose causal diagram of the equatorial plane of the 
Kerr spacetime exterior to the future and past horizons. Each
point corresponds to a circle. $i^{\pm}$ are future and past timelike infinity,
$i^0$ is spatial infinity, B is the bifurcation surface, and $\cal H$ is
the future horizon. The dashed lines indicate Boyer-Linquist and
Doran time slices as indicated. The two curves from $i^-$ to $i^+$
indicate the photon orbit (closer to $\cal H$) and the ISCO. The outer boundary of 
the gray region indicates the surface of matter that collapsed to form
the black hole, so the gray region is not present in an astrophysical spacetime.
In the extremal limit, the orbits both converge to the left edge of the diagram, and
the proper distance to $B$ along the Boyer-Lindquist time slice diverges, 
whereas the distance to $\cal H$ along the Doran time slice remains finite.
\label{CPdiagram} }
\end{figure}

On the other hand, the Kerr metric can of course also be 
described using time slices that cross the future horizon, like the 
Doran time slice in Fig.\ 1.
For a rotating black hole that forms from gravitational collapse, 
only slices that cross the future horizon can remain in the stationary
portion of the spacetime. On such a slice, points with 
the same value of the radial and angular coordinates are certainly 
at the same location, so the extremal limit of the photon orbit 
and ISCO must in fact lie on the future horizon. 

In occasional articles and private conversations
I have encountered confusion generated by the result of Ref.~\cite{Bardeen:1972fi}.
Some of the confusion may arise from the way things are 
described in Ref.~\cite{Bardeen:1972fi}. It is written there in section II that in the extremal limit,
``it appears that the photon, marginally bound, and marginally stable orbits are coincident
with the horizon. Appearances are deceptive!" It then goes on to say that in this 
limit the orbits ``remain separated in proper radial distance" yet, after mentioning
that some of these distances are infinite, adds parenthetically ``The infinities are not physically
important; an infalling particle follows a timelike curve, while the infinite distances are in a 
spacelike direction." This seems to clear things up but 
then in section III, after observing that the velocity of the ISCO 
measured relative to the locally non-rotating frame  
goes to $1/2$ and not $1$ in the extremal limit, it is stated: 
``The point once again is that for $a=M$, the marginally stable orbit 
and the photon orbit are distinct." 
The aim of this note is to elucidate the
compatibility of the two viewpoints, and to stress that on a horizon-crossing
time slice the orbits in the limit are {\it not} distinct,
and indeed coincide with 
the horizon generators.

\section{Boyer-Lindquist and Doran coordinates}
\label{coordinates}

The Kerr black hole spacetime has both time translation symmetry
and axial rotation symmetry. We will describe it here using both the
Boyer-Lindquist and the Doran\cite{Doran:1999gb}
coordinate systems. 
These have in common two coordinates, called $r$ and $\theta$,
which are constant in the symmetry
directions. The remaining two coordinates will be called 
$(t,\phi)$ for BL and $(\tb,\phib)$ for Doran.
The line element 
in these two coordinate systems  takes the form
\bea
ds^2&=&-(1-\a^2)\, dt^2
+\a^{-2}\b^2(1-\b^2)^{-1}\, dr^2
-2 \a^2 a\sin^2\!\theta\, dt d\phi\nonumber\\
&&\qquad\qquad+\r^2\, d\theta^2
+(r^2+a^2+\a^2 a^2 \sin^2\!\theta)\sin^2\!\theta\, d \phi^2\label{BL}\\
\nn\\
&=&-d\tb^2 
+ \left(\a^{-1}\b\,  dr 
+ \a( d\tb - a\sin^2\!\theta \, d\phib) \right)^2 \nn \\
&&\qquad\qquad + \rho^2\, d\theta^2 + (r^2 +a^2) \sin^2\!\theta \, d\phib^2,
\label{Doran}
\eea
with 
\beq\label{alpharho}
\a^2=\frac{2Mr}{\r^2},
\qquad
\b^2=\frac{2Mr}{r^2+a^2},
\qquad \r^2 = r^2 + a^2\cos^2\!\theta,
\eeq
where $\a$ and $\beta$ are the positive roots.
The constants $M$ and $a$ are the mass and spin parameter
(angular momentum divided by mass) of the black hole,
and we use units with $G=c=1$.

The ``time" coordinates $t$ and $\tb$
differ from each other only by the addition of a function
of $r$, as do the azimuthal angle coordinates
$\phi$ and $\phib$:
\beq\label{tphi}
d\tb= dt+{\frac{\b}{1-\b^2}}\, dr, 
\qquad
d\phib=d\phi+\frac{\b}{1-\b^2}\frac{a}{r^2+a^2}\, dr
\eeq
Note that, since the coordinates differ only by functions of $r$, 
the metric coefficients of all terms not involving $dr$ are
equal for the two coordinate systems.  In the non-rotating
case $a=0$, BL becomes Schwarzschild 
and Doran becomes Painlev\'e-Gullstrand.

In BL coordinates the metric has just one off-diagonal ($dt d\phi$)
term,
and the constant $t$ surfaces are orthogonal to 
stationary, zero angular momentum observers.
In Doran coordinates,
the curves  $d\th=d\phib=\b dr+\a^2d\tb=0$ are infalling
timelike geodesics with zero angular momentum, unit energy
(at rest at infinity), and proper
time $d\tb$. The surfaces
of constant $\tb$ are orthogonal to
these geodesics.
The event horizon of the black hole is located where
$\b=1$, i.e. at
\beq
r_h=M+\sqrt{M^2-a^2}.
\eeq
Doran coordinates are regular at the horizon,
but the BL $t$ and $\phi$ coordinates both 
run to infinity there at constant $\tb$ and $\phib$, 
as can be seen from the 
coordinate transformation (\ref{tphi}). 
In the extremal case, $a=M$, 
the horizon is located at $r_h=M$.
From here on we adopt coordinates with 
$M=1$.

We are interested in the
geometries of the equatorial 
constant time slices, which are given by 
\bea
dl^2_{\rm BL}&=&
\frac{r^2}{(r-1-\e)(r-1+\e)}\, dr^2+\left(r^2 +a^2 +\frac{2Ma^2}{r}\right)d\phi^2\label{BLslice}\\
dl^2_{\rm Doran}&=&
\frac{r^2}{r^2+a^2}\,  dr^2 -\sqrt{\frac{8r}{r^2+a^2}}\, dr\, d\phib+\left(r^2+a^2+\frac{2Ma^2}{r}\right) d\phib^2,
\eea
where $\e=r_h-1=\sqrt{1-a^2}$ is the deviation of the horizon radius from $r=1$. 
%
%
In the extremal limit we have $\e\rightarrow0$, and 
on a BL time slice 
the radial distance to the bifurcation surface diverges as $\ln \e$.\footnote{This 
refers to the distance
orthogonal to the circles of constant $r$, at constant
$\phi$. Note 
that the two angles
$\phi$ and $\phib$ wrap infinitely relative to each other as the
horizon is approached (\ref{tphi}). 
The distances to the horizon
on a BL slice at constant $\phib$ and on a Doran slice at constant $\phi$ both 
diverge logarithmically in the non-extremal case and linearly in the extremal case.}

\section{Location of orbits in the extremal limit}
Let us now see how the photon orbit and ISCO can behave as described
above. 
First, it is clear that on the Doran slice, since the limiting metric
is regular in the $r$ coordinate at the horizon, the fact that the
radial coordinate of these orbits tends to $r_h$ means that they must in fact 
coincide with the horizon. This can also be
characterized by the proper time interval along an infalling,
zero angular momentum geodesic as these radii are crossed.
On such a geodesic the proper time $d\tb$ is related to $dr$ by 
$d\tb = (\b/\a^2)dr$ [{\it cf}. (\ref{Doran})], which at the horizon 
in the extremal limit becomes $d\tb = dr/2$. The infalling proper time from the
photon orbit or ISCO to the horizon thus goes to zero. 
This illustrates and verifies the fact that these
orbits both coalesce to the horizon generators. 

The photon orbit
is lightlike, so it is not surprising that it can converge to the horizon
generators, but the ISCO is timelike. Evidently the 4-velocity
of the ISCO must diverge as a vector in this limit. In fact, using the
results of Ref.~\cite{Bardeen:1972fi} with help from Mathematica,
one finds that the Doran components of the ISCO 4-velocity are given by 
\bea
\frac{d\tb}{ds}&=&\frac{2^{5/3}}{\sqrt{3}}\e^{-2/3}-\frac{\sqrt{3}}{2}+O(\e^{2/3})\label{4velt}\\
\frac{d\phib}{ds}&=&\frac{2^{2/3}}{\sqrt{3}}\e^{-2/3}-\frac{3\sqrt{3}}{4}+O(\e^{2/3}).\label{4velphi}
\eea
(The BL components are the same, since the orbit lies at constant $r$.)
Despite the divergence in the 4-velocity, the energy and angular momentum per unit
mass tend to the finite values $1/\sqrt{3}$ and $2/\sqrt{3}$ respectively.\footnote{To obtain 
the $O(\e^0)$ terms in (\ref{4velt},\ref{4velphi}) it is necessary to expand the ISCO radius
out to $O(\e^{8/3})$. The finite terms are needed if the
energy and angular momentum are computed directly using the 4-velocity.}
The limiting frequency shift factor of a photon emitted from the ISCO in the transverse direction,
i.e.\ orthogonal to the orbit, is $(dt/ds)^{-1}\sim\e^{2/3}$. By this redshift a distant observer 
could in principle infer that the orbit coincides with the horizon in the extremal limit.\footnote{The convergence
is slow however: a spin parameter $a=0.998$ (which corresponds to $\e\sim0.06$)
yields  horizon radius $\sim 1.06$,
photon orbit radius $\sim 1.07$, ISCO radius $\sim 1.24$, and ISCO redshift factor $\sim 0.09$.}
As mentioned in the introduction,
the limiting velocity of the ISCO relative to the 
``locally non-rotating frame" is $1/2$~\cite{Bardeen:1972fi}.
The 4-velocity of that frame also stretches to an infinite vector and approaches 
the horizon generator, so the fact that this relative velocity is less than 1 is not inconsistent
with the ISCO approaching the speed of light. 

Now let us examine the distance between the orbits on a BL slice.
The perpendicular distance between circles of two radii 
$r_{1,2}=1+\d_{1,2}$ on the BL slice (\ref{BLslice})
is given by
\beq
\D l 
=\left(\sqrt{x^2-\e^2}+\ln\left[2(x+\sqrt{x^2-\e^2}\right] \right)^{\d_2}_{\d_1},
\eeq
where $x=r-1$.
If $\d_2$ remains finite and $\d_1,\e\rightarrow0$, then the distance
$\D l$ diverges as $\ln \d_1$. If both $\d_1,\d_2\rightarrow0$, 
then the distance depends on precisely how they 
approach zero. 
Suppose that 
$\d_1=c_1\e^{p_1}$ and $\d_2=c_2\e^{p_2}$, 
where $c_{1,2}>0$ and $0< p_2\le p_1\le1$
are constants. Then we have
\beq
\lim_{\e\rightarrow0}\D l 
=\left\{ \begin{array}{ll}
(p_2-p_1)\ln\e&p_2<p_1\\
\ln(c_2/c_1)&p_2=p_1<1\\
\ln\left[\frac{c_2+\sqrt{c_2^2-1}}{c_1+\sqrt{c_1^2-1}}\right] &p_2=p_1=1
\end{array}
\right.
\eeq

In particular, according to Ref~\cite{Bardeen:1972fi} 
the radial coordinate for the photon orbit 
is $r_{\rm ph}\approx1 +(2/\sqrt{3})\e$, and for the ISCO it is
$r_{\rm isco}\approx 1+2^{1/3}\e^{2/3}$. Thus the limiting distance from the
photon orbit to the bifurcation surface is $\D l(r_h,r_{\rm ph})=(1/2)\ln{3}$, 
the distance from the ISCO to the bifurcation surface diverges as 
$\D l(r_h,r_{\rm isco})\sim -(1/3)\ln\e$, and the distance from the ISCO
to the ergosurface at $r=2$ diverges as 
$\D l(r_{\rm isco},r_{\rm ergo})\sim -(2/3)\ln\e$.

In conclusion,  the orbits both coincide and do not coincide,
depending on where they are examined. This is possible since 
in the extremal limit there is
an infinite stretching of the spacetime region outside the horizon 
along a constant BL time slice\footnote{The extremal metric in the neighborhood 
of the bifurcation surface is written in Ref.~\cite{Bardeen:1999px} using 
rescaled time and radius coordinates 
$T=\lambda t$, $R=(r-M)/\lambda$, and shifted angle $\Phi=\phi-t/2M$,
and taking the limit $\lambda\rightarrow0$.
For that metric one can show that there are marginally stable orbits 
at all radii, all with the same energy and angular momentum
as that of the extremal ISCO.}, 
and no such stretching on
a horizon-crossing time slice.

\section*{Acknowledgements}
I thank E.~Barausse, A.~Buonanno, and 
E.~Poisson for useful comments on a draft of this paper.
This work was supported in part by NSF grant PHY-0903572.


\begin{thebibliography}{10}

\bibitem{Bardeen:1972fi}
  J.~M.~Bardeen, W.~H.~Press, S.~A.~Teukolsky,
  ``Rotating black holes: Locally nonrotating frames, energy extraction, and scalar synchrotron radiation,''
  Astrophys.\ J.\  {\bf 178}, 347 (1972).
  
\bibitem{Doran:1999gb}
  C.~Doran,
  ``A New form of the Kerr solution,''
  Phys.\ Rev.\  {\bf D61}, 067503 (2000)
  [gr-qc/9910099].
  
\bibitem{Bardeen:1999px}
  J.~M.~Bardeen and G.~T.~Horowitz,
  ``The Extreme Kerr throat geometry: A Vacuum analog of AdS$_2$ $\times$ S$^2$,''
  Phys.\ Rev.\  D {\bf 60}, 104030 (1999)
  [arXiv:hep-th/9905099].

\end{thebibliography}
\end{document}